\documentclass[twocolumn,showpacs,superscriptaddress,amsmath,amssymb,aps,pra]{revtex4-2}
\usepackage{graphicx}
\usepackage{CJKutf8}
\usepackage{dcolumn}
\usepackage{amsmath,bm}
\usepackage{epstopdf}
\usepackage[mathlines]{lineno}
\usepackage{color}
\usepackage[colorlinks=true,linkcolor=blue,citecolor=magenta,urlcolor=cyan]{hyperref}

\usepackage{tikz,xcolor,hyperref}

\definecolor{lime}{HTML}{A6CE39}
\DeclareRobustCommand{\orcidicon}{%
	\begin{tikzpicture}
	\draw[lime, fill=lime] (0,0) 
	circle [radius=0.16] 
	node[white] {{\fontfamily{qag}\selectfont \tiny ID}};
	\draw[white, fill=white] (-0.068,0.105) 
	circle [radius=0.007];
	\end{tikzpicture}
	\hspace{-2mm}
}

\foreach \x in {A, ..., Z}{%
	\expandafter\xdef\csname orcid\x\endcsname{\noexpand\href{https://orcid.org/\csname orcidauthor\x\endcsname}{\noexpand\orcidicon}}
}

\begin{document}
\title{Low-energy subband wave-functions and effective $g$-factor of one-dimensional hole gas}
\author{Rui\! Li~(\begin{CJK}{UTF8}{gbsn}李睿\end{CJK})\orcidA{}}
\email{ruili@ysu.edu.cn}
\affiliation{Key Laboratory for Microstructural Material Physics of Hebei Province, School of Science, Yanshan University, Qinhuangdao 066004, China}
\begin{abstract}
One-dimensional hole gas confined in a cylindrical Ge nanowire has potential applications in quantum information technologies. Here, we analytically study the low-energy properties of this one-dimensional hole gas. The subbands of the hole gas are two-fold degenerate. The low-energy subband wave-functions are obtained exactly, and the degenerate pairs are related to each other via a combination of the time-reversal and the spin-rotation transformations. In evaluating the effective $g$-factor of these low-energy subbands, the orbital effects of the magnetic field are shown to contribute as strongly as the Zeeman term. Also, near the center of the $k_{z}$ space, there is a sharp dip or a sharp peak in the effective $g$-factor. At the site $k_{z}=0$, the longitudinal $g$-factor $g_{l}$ is much less than the transverse $g$-factor $g_{t}$ for the lowest subband, while away from the site $k_{z}=0$, $g_{l}$ can be comparable to $g_{t}$.
\end{abstract}
\date{\today}
\maketitle

\section{Introduction}
Quantum computation based on the electron spins in semiconductor quantum dots has obtained a series of important advances in recent decades~\cite{loss1998quantum,petta2005coherent,hanson2007spins,vandersypen2019}. A strong single-spin manipulation can be achieved using the electric-dipole spin resonance technique, an interesting phenomena induced by the strong spin-orbit coupling in semiconductor quantum dot~\cite{Nowack1430,golovach2006electric,trif2008spin,lirui2013controlling,nadj2010spin}. Two-spin manipulation can be implemented using the exchange interaction of two electrons in a semiconductor double quantum dot~\cite{burkard1999coupled,hu2000hilbert}. However, like many other qubit candidates, the quantum dot electron spin also suffers from the decoherence problem. One  main spin dephasing comes from the lattice nuclear spins, who couple to the electron spin via the hyperfine interaction~\cite{witzel2006quantum,yao2006theory,cywinski2009electron}. Another spin dephasing comes from the charge noise, a longitudinal spin-charge interaction can be mediated by the interplay between the spin-orbit coupling and the asymmetrical quantum dot confining potential~\cite{yoneda2018,lirui2018a,Li2020charge}.

Recently, extensive studies have focused on the hole spins in semiconductor quantum dots~\cite{PhysRevLett.101.186802,PhysRevLett.112.216806,PhysRevB.94.041411,Watzinger:2018aa}. The hole spin has a few good properties, such that it is an excellent qubit candidate as well as the electron spin. First, due to the p-type character of the valence band of the semiconductors~\cite{PhysRev.97.869,PhysRev.102.1030,winkler2003spin}, the hyperfine interaction between the hole spin and the lattice nuclear spins is suppressed. Second, the intrinsic spin-orbit coupling for hole is very large~\cite{PhysRev.97.869,PhysRev.102.1030,PhysRevB.63.235302,PhysRevB.64.085329}, such that hole spin is also manipulable by an external electric field~\cite{PhysRevLett.98.097202,PhysRevB.99.115317,PhysRevB.95.195316,wang2020ultrafast}. As is well known, the low-energy physics of the electron near the bottom of the conductor band is well described by a parabolic band dispersion~\cite{winkler2003spin}. On the other hand, the low-energy physics of the hole near the top of the valence band is described by the 4-bands Luttinger-Kohn Hamiltonian~\cite{PhysRev.97.869,PhysRev.102.1030,PhysRevB.52.11132,PhysRevB.71.075308,PhysRevLett.95.076805}. Although the 4-bands Luttinger-Kohn Hamiltonian seems more complicated, it actually leads to two branches of parabolic dispersion can be labelled with heavy hole and light hole, respectively~\cite{WU201061}.

Semiconductor Ge gets a special attention in the studies of the hole spin~\cite{Scappucci:2020aa}. From the aspects of experimental studies, this is because Ge can be engineered into free of nuclear spins, and it has a relative large $g$-factor~\cite{PhysRevB.4.3460}. Also, from the aspects of theoretical studies, we can conveniently make the spherical approximation to the Luttinger-Kohn Hamiltonian~\cite{PhysRev.102.1030}. For hole spins confined in a planar Ge quantum dot~\cite{Hendrickx:2020ab,Hendrickx:2020aa}, the electric manipulation~\cite{PhysRevB.103.125201} and the dephasing~\cite{Wang:2021wc} of the hole spin were studied. In recent years, hole spins confined in quasi one-dimensional (1D) Ge nanowire quantum dot are getting increasing attention. Quasi-1D hole gas can be realized in a Ge hut wire~\cite{Watzinger:2016aa,Li:2018aa,Gao2020AM} or in a Ge/Si core-shell nanowire~\cite{PhysRevB.95.155416,Froning:2021aa}. A quasi-1D quantum dot can be achieved by placing proper metallic gates bellow the 1D hole gas~\cite{PhysRevLett.101.186802,PhysRevLett.112.216806,PhysRevB.87.161305}. We are motivated to study the hole properties in these simple 1D systems. Our first step is to understand the low-energy subband dispersions of the 1D hole gas, and especially the effective $g$-factor of these subbands.

In this paper, we follow the method introduced in the seminal paper~\cite{PhysRevB.42.3690} to study the low-energy properties of the hole gas confined in a cylindrical Ge nanowire. We first calculate the exact low-energy subband wave-functions of the hole gas, even for those wave vectors are not at the center of the $k_{z}$ space, i.e., $k_{z}\neq0$. The subbands are two-fold degenerate. Via exploring the symmetries of the hole Hamiltonian, the two degenerate subband wave-functions are related to each other via a combination of the time-reversal transformation and the spin-rotation transformation. We then consider the effects of the external magnetic fields which are applied to the hole gas both longitudinally and transversely. In the contributions to the subband splitting induced by the external magnetic field, we find the orbital effects of the magnetic field are as important as the Zeeman term. There is a sharp dip or a sharp peak at the site $k_{z}=0$ when the subband $g$-factor is plotted as a function of $k_{z}$. This is due to the reason $k_{z}=0$ is an energy anticrossing site in the subband dispersions.

The paper is organized as follows. In Sec.~\ref{Sec_II}, we give the model of the 1D hole gas. In Sec.~\ref{Sec_III}, we calcu­late exactly the low-energy subbands and the corresponding subband­ wave-functions. In Sec.~\ref{Sec_IV}, we study the longitudinal subband $g$-factor. In Sec.~\ref{Sec_V}, we study the transverse subband $g$-factor. In Sec.~\ref{Sec_VI}, we give a short discussion on the hole spin qubit in quantum dot. At last, we conclude in Sec.~\ref{Sec_VII}.
 
\begin{figure}
\includegraphics{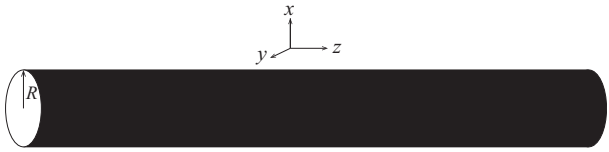}
\caption{\label{Fig_nanowire}A cylindrical Ge nanowire of radius $R$ in the hole regime is under investigation.}
\end{figure}

\section{\label{Sec_II}The model and the associated symmetries}
The system we are interested in is explicitly shown in Fig.~\ref{Fig_nanowire}. A hole is allowed to move in a cylindrical Ge nanowire, the wire can be grown experimentally via chemical vapor deposition method~\cite{PhysRevLett.101.186802,PhysRevLett.112.216806}. We use the Luttinger-Kohn Hamiltonian~\cite{PhysRev.97.869,PhysRevB.52.11132,WU201061} in the spherical approximation~\cite{PhysRev.102.1030} to describe the kinetic energy of the hole in the wire, this is indeed a good approximation for semiconductor Ge. In order to think of the energy levels in the more usual way, we put an overall minus sign to the hole Hamiltonian~\cite{PhysRev.97.869}. The whole Hamiltonian of the hole under study reads~\cite{PhysRevB.42.3690,PhysRevB.84.195314} (the real hole Hamiltonian should be interpreted as $-H_{0}$)
\begin{equation}
H_{0}=\frac{1}{2m_{e}}\left[\left(\gamma_{1}+\frac{5}{2}\gamma_{s}\right)\textbf{p}^{2}-2\gamma_{s}(\textbf{p}\cdot\textbf{J})^{2}\right]+V(r),\label{Eq_model}
\end{equation}
where $m_{e}$ is the bare electron mass, $\gamma_{1}=13.35$ and $\gamma_{s}=(2\gamma_{2}+3\gamma_{3})/5=5.11$ are Luttinger parameters~\cite{PhysRevB.4.3460}, ${\bf p}=-i\hbar\nabla$ is the momentum operator, ${\bf J}=(J_{x}, J_{y}, J_{z})$ is a spin-$3/2$ vector operator, with the explicit matrix forms of $J_{x,y,z}$ being given in Appendix~\ref{Appendix_A}, and $V(r)$ is the transverse ($xy$ plane) confining potential of the hole. In this paper, we use a cylindrical hard wall to model this potential~\cite{PhysRevB.79.155323,PhysRevB.84.195314}
\begin{equation}
V(r)=\left\{\begin{array}{cc}0,~&~r<R,\\
\infty,~&~r>R,\end{array}\right.\label{Eq_potential}
\end{equation}
where $R$ is the radius of the Ge nanowire. Note that the experimentally achievable $R$ is about $5\sim10$ nm~\cite{PhysRevLett.101.186802,PhysRevLett.112.216806}.

Let us analyze the symmetries associated with our model (\ref{Eq_model}). The symmetry of the model not only can simplify the corresponding calculations in the following, but also can give a simple physical picture of the induced subband dispersions. First, it is easy to know $p_{z}=\hbar\,k_{z}$ is a conserved quantity in Hamiltonian (\ref{Eq_model}), such that the energy eigenvalues of which can be written as a 1D band dispersion $E_{n}(k_{z})$, where $n$ is the subband index. Second, the Hamiltonian (\ref{Eq_model}) is time-reversal invariant. The time-reversal operator can be written as
\begin{equation}
T=\Gamma_{1}\Gamma_{3}K,
\end{equation}
where $K$ is the usual complex conjugate operator, and $\Gamma_{1,3}$ are two $4\times4$ matrices~\cite{PhysRevLett.91.186402}. Note that the detailed expressions of $\Gamma_{1,3}$ are given in Appendix {\ref{Appendix_A}}. One can verify that $T^{2}=-1$, $T{\bf J}T^{-1}=-{\bf J}$, and $T{\bf p}T^{-1}=-{\bf p}$. It follows that our Hamiltonian (\ref{Eq_model}) is time-reversal invariant $TH_{0}T^{-1}=H_{0}$. Therefore, we have the Kramer's degeneracy 
\begin{equation}
E_{n}(k_{z},\Uparrow)=E_{n}(-k_{z},\Downarrow),\label{Eq_symmetry1}
\end{equation}
where $\Uparrow$ and $\Downarrow$ are introduced to distinguish the spin orientations in the Kramer's doublets. Third, we introduce a spin-rotation operator $U(\theta)=e^{i\theta\,J_{z}}$.
Applying a unitary transformation of $\pi$ rotation to the Hamiltonian (\ref{Eq_model}), we have $U(\pi)H_{p_{z}}U^{\dagger}(\pi)=H_{-p_{z}}$. This equation gives rise to
\begin{equation}
E_{n}(k_{z})=E_{n}(-k_{z}).\label{Eq_symmetry2}
\end{equation}
Combining Eqs.~(\ref{Eq_symmetry1}) and (\ref{Eq_symmetry2}), we have
\begin{equation}
E_{n}(k_{z},\Uparrow)=E_{n}(k_{z},\Downarrow).
\end{equation}
Hence, for a given wave vector $k_{z}$ along the wire direction, we have the spin degeneracy in the subband dispersions.

\section{\label{Sec_III}Subband dispersions and subband wave-functions}
Without the hard-wall potential (\ref{Eq_potential}) coming from the transverse boundary of the nanowire, the bulk hole system is described by the Luttinger-Kohn Hamiltonian in the spherical approximation~\cite{PhysRev.102.1030}
\begin{equation}
H_{LK}=\frac{1}{2m_{e}}\left[\left(\gamma_{1}+\frac{5}{2}\gamma_{s}\right)\textbf{p}^{2}-2\gamma_{s}(\textbf{p}\cdot\textbf{J})^{2}\right].\label{Eq_LK}
\end{equation}
It is easy to obtain the hole bulk spectrum by solving the eigenvalue equation of Hamiltonian~(\ref{Eq_LK}) in momentum space, one gets~\cite{WU201061}
\begin{equation}
E=(\gamma_{1}\pm2\gamma_{s})\frac{\hbar^{2}(\mu^{2}+k^{2}_{z})}{2m_{e}},
\end{equation}
where $\hbar^{2}\mu^{2}=p^{2}_{x}+p^{2}_{y}$. The bulk wave-functions are also very important in deriving the hole subbands in the nanowire, such that we need to obtain both the bulk spectrum and the corresponding bulk wave-functions using cylindrical coordinate system in order to respect the cylindrical boundary condition (\ref{Eq_potential}) of the nanowire.

We study the Luttinger-Kohn Hamiltonian in cylindrical coordinate system by writing $x=r\cos\varphi$, $y=r\sin\varphi$, $z=z$. The $z$-component of the total angular momentum $F_{z}=-i\partial_{\varphi}+J_{z}$ is a conserved quantity, i.e., $[H_{LK},F_{z}]=[H_{0},F_{z}]=0$, it follows that we can classify the wave-functions using $F_{z}$~\cite{PhysRevB.42.3690,PhysRevB.79.155323,PhysRevB.84.195314}. The detailed derivations of the bulk spectrum and the corresponding bulk wave-functions were already given in Ref.~\cite{PhysRevB.42.3690}. We summarize these results in Appendix~\ref{Appendix_B}, which would be very helpful in the following calculations. Note that we have chosen a different representation for spin operators $J_{x,y,z}$, such that some expressions given here are a little different from that of Ref.~\cite{PhysRevB.42.3690}.

We proceed to derive the subband dispersions in the nanowire by taking into account the cylindrical boundary condition. The eigenfunction of Hamiltonian (\ref{Eq_model}) can be written as a linear superposition of the four bulk wave-functions, i.e.,
\begin{eqnarray}
\Psi(r,\varphi,z)&\equiv&\left(\begin{array}{c}\Psi_{1}(r)e^{i(m-1)\varphi}\\\Psi_{2}(r)e^{im\varphi}\\\Psi_{3}(r)e^{i(m+1)\varphi}\\\Psi_{4}(r)e^{i(m+2)\varphi}\end{array}\right)e^{ik_{z}z},\label{Eq_eigenfunc}
\end{eqnarray}
where
\begin{eqnarray}
\Psi_{1}(r)&=&c_{1}\frac{2ik_{z}}{\mu_{1}}J_{m-1}(\mu_{1}\,r)+c_{2}\sqrt{3}J_{m-1}(\mu_{1}\,r)\nonumber\\
&&+c_{3}\frac{2ik_{z}}{\mu_{2}}J_{m-1}(\mu_{2}\,r)-c_{4}\frac{4k^{2}_{z}+\mu^{2}_{2}}{\sqrt{3}\mu^{2}_{2}}J_{m-1}(\mu_{2}\,r),\nonumber\\
\Psi_{2}(r)&=&c_{1}\frac{4k^{2}_{z}+\mu^{2}_{1}}{\sqrt{3}\mu^{2}_{1}}J_{m}(\mu_{1}\,r)-c_{2}\frac{2ik_{z}}{\mu_{1}}J_{m}(\mu_{1}\,r)\nonumber\\
&&-c_{3}\sqrt{3}J_{m}(\mu_{2}\,r)-c_{4}\frac{2ik_{z}}{\mu_{2}}J_{m}(\mu_{2}\,r),\nonumber\\
\Psi_{3}(r)&=&c_{2}J_{m+1}(\mu_{1}\,r)+c_{4}J_{m+1}(\mu_{2}\,r),\nonumber\\
\Psi_{4}(r)&=&c_{1}J_{m+2}(\mu_{1}\,r)+c_{3}J_{m+2}(\mu_{2}\,r),\label{eq_wavefunction}
\end{eqnarray}
with $J_{m}(\mu\,r)$ being the $m$-order Bessel function~\cite{PhysRevB.42.3690}, $\mu_{1,2}= \sqrt{2m_{e}E/[(\gamma_{1}\pm2\gamma_{s})\hbar^{2}]-k^{2}_{z}}$, and $c_{1,2,3,4}$ being the expansion coefficients to be determined. It should be noted that for eigenfunction (\ref{Eq_eigenfunc}), the total angular momentum has the value $F_{z}=m+1/2$, where $m=0,\pm1,\pm2,\ldots$. Subjecting the eigenfunction (\ref{Eq_eigenfunc}) to the hard wall boundary condition $\Psi(R,\varphi,z)=0$, one obtains a transcendental equation of $E$, the solution of which gives all the allowed energies of the hole in the nanowire (for details see Appendix~\ref{Appendix_B}).
 
\begin{figure}
\includegraphics{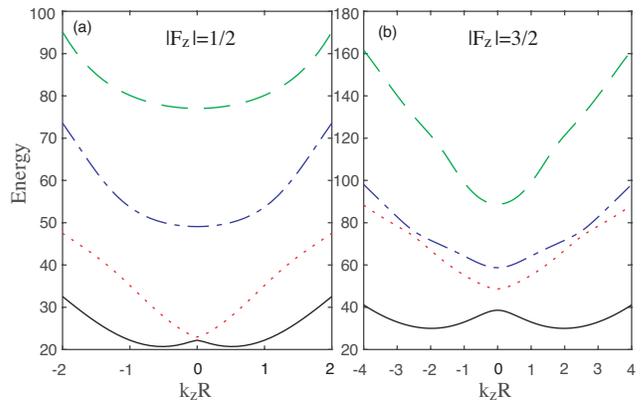}
\caption{\label{Fig_subband}Subband dispersions of the 1D hole gas. The energy is in unit of $\hbar^{2}/(m_{e}R^{2})$. (a) The lowest four subbands for total angular momentum $|F_{z}|=1/2$. (b) The lowest four subbands for total angular momentum $|F_{z}|=3/2$. }
\end{figure}

For a given wave vector $k_{z}$ along the nanowire, we can obtain a series of $E_{n}$ via solving Eq.~(\ref{eq_transcendental}) in Appendix~\ref{Appendix_B}. We show the obtained low-energy subband dispersions in Fig.~\ref{Fig_subband}. Figures~\ref{Fig_subband}(a) and (b) give the results for total angular momentums $|F_{z}|=1/2$ and $|F_{z}|=3/2$, respectively. As can be seen from Figs.~\ref{Fig_subband}(a) and (b), the band minimum is not at the center, i.e., at $k_{z}=0$, of the $k_{z}$ space~\cite{PhysRevB.84.195314}. Also, the two lowest subbands [shown in Fig.~\ref{Fig_subband}(a)] are very close to each other, and they are well separated from the other higher subbands for relative small $k_{z}$. In particular, these two lowest subbands of the hole gas are very similar to that of a 1D electron gas with both strong Rashba SOC and external magnetic field~\cite{PhysRevB.84.195314}, e.g., described by the Hamiltonian $H_{c}=p^{2}/(2m_{e})+\alpha\sigma^{z}p+(g_{e}\mu_{B}B/2)\sigma^{x}$~\cite{lirui2018energy,lirui2018the}. The key difference is that there is a spin degeneracy for the hole case, as is illustrated in Sec.~\ref{Sec_II}. This similarity has also inspired a series of studies trying to realize a strong spin-orbital coupled 1D hole gas via breaking the spin degeneracy~\cite{PhysRevB.84.195314,PhysRevB.87.161305}.

\begin{figure}
\includegraphics{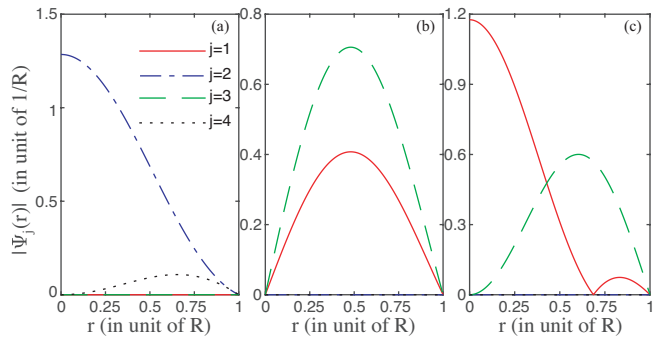}
\caption{\label{Fig_WF00}The distribution of the subband wave-function at $k_{z}R=0$ along the transverse direction of the nanowire. (a) The result for the lowest subband with total angular momentum $F_{z}=1/2$. (b) The result for the second lowest subband with total angular momentum $F_{z}=1/2$. (c) The result for the lowest subband with total angular momentum $F_{z}=3/2$.}
\end{figure}

\begin{figure}
\includegraphics{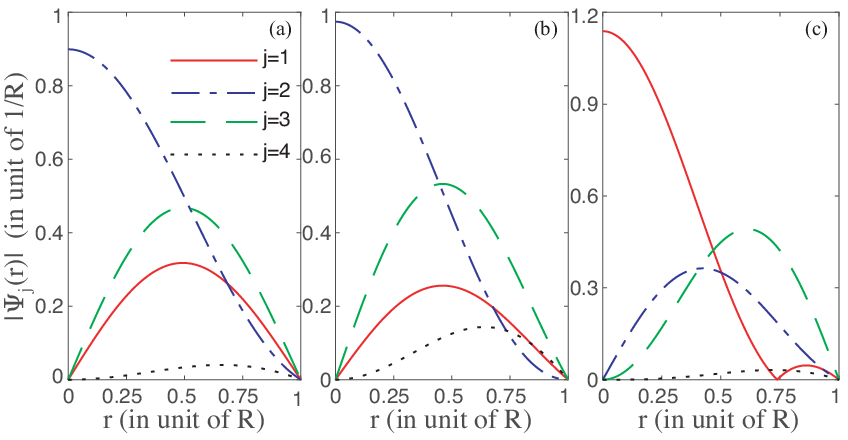}
\caption{\label{Fig_WF05}The distribution of the subband wave-function at $k_{z}R=0.5$ along the transverse direction of the nanowire. (a) The result for the lowest subband with total angular momentum $F_{z}=1/2$. (b) The result for the second lowest subband with total angular momentum $F_{z}=1/2$. (c) The result for the lowest subband with total angular momentum $F_{z}=3/2$.}
\end{figure}

\begin{figure}
\includegraphics{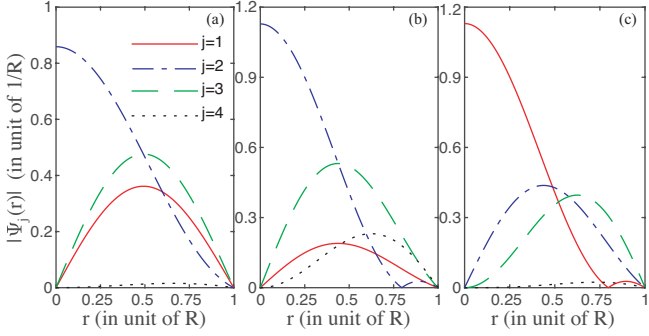}
\caption{\label{Fig_WF09}The distribution of the subband wave-function at $k_{z}R=0.9$ along the transverse direction of the nanowire. (a) The result for the lowest subband with total angular momentum $F_{z}=1/2$. (b) The result for the second lowest subband with total angular momentum $F_{z}=1/2$. (c) The result for the lowest subband with total angular momentum $F_{z}=3/2$.}
\end{figure}

Once an energy eigenvalue, say the $n$-th eigenvalue $E_{n}(k_{z})$ for a given wave vector $k_{z}$, is obtained, we can evaluate the corresponding expansion coefficients $c^{(n)}_{1,2,3,4}$ via solving the equation array (\ref{Eq_coeffeicents}) (in combination with the normalization condition). Once the coefficients $c^{(n)}_{1,2,3,4}$ are obtained, the eigenfunction corresponds to  a given energy eigenvalue $E_{n}(k_{z})$ is also known from Eq.~(\ref{eq_wavefunction}). In Figs.~\ref{Fig_WF00}, \ref{Fig_WF05}, and \ref{Fig_WF09}, we show the distributions of the eigenfunctions along the transverse direction of nanowire for special wave vectors $k_{z}R=0$, $0.5$, and $0.9$, respectively. We have shown the lowest three subband wave-functions at these special $k_{z}$ values, i.e., the two lowest subbands with total angular momentum $F_{z}=1/2$ shown in Fig.~\ref{Fig_subband}(a) and the lowest subband with total angular momentum $F_{z}=3/2$ shown in Fig.~\ref{Fig_subband}(b). 

As is shown in Fig.~\ref{Fig_WF00}, two components of the subband wave-function (\ref{Eq_eigenfunc}) at the wave vector $k_{z}R=0$ are zero. While at the site $k_{z}R\neq0$, e.g., $k_{z}R=0.5$ shown in Fig.~\ref{Fig_WF05} or $k_{z}R=0.9$ shown in Fig.~\ref{Fig_WF09}, all the four components of the subband wave-function have finite values. Actually, the calculations at the site $k_{z}R=0$ are much simpler than that at the site $k_{z}R\neq0$. The transcendental equation of $E$, given by Eq.~(\ref{eq_transcendental}), has a very simple form when $k_{z}R=0$~\cite{PhysRevB.76.073313,PhysRevB.84.195314}. This is due to the reason the Luttinger-Kohn Hamiltonian can be block diagonalized at this special site~\cite{PhysRevB.79.155323}. Note that the lowest band dispersion given in Fig.~\ref{Fig_subband}(a) has its minimum nearly at $|k_{z}R|=0.5$, and there is an anticossing between the second and third bands in Fig.~\ref{Fig_subband}(a) nearly at the site $|k_{z}R|=0.9$. Note that here we only show the distribution of the components of the eigenfunction instead of the total probability density distribution of the eigenfunction~\cite{SAIDI2020136872}. This is because the distribution of the components can give us more intuitive information about the eigenfunction.

Although for a given wave vector $k_{z}$, the two degenerate eigenfunctions can be obtained via choosing proper values of the total angular momentum $F_{z}$, e.g., $F_{z}=m+1/2$ and $F_{z}=-(m+1/2)$ give the same energy eigenvalue. However, there is a simple way to obtain the degenerate counterpart of a given eigenfunction. Actually, from the symmetries discussed in Sec.~\ref{Sec_II}, we also conclude that the following two states have the same energy eigenvalue
\begin{equation}
\left(\begin{array}{c}\Psi_{1}(r)e^{i(m-1)\varphi}\\\Psi_{2}(r)e^{im\varphi}\\\Psi_{3}(r)e^{i(m+1)\varphi}\\\Psi_{4}(r)e^{i(m+2)\varphi}\end{array}\right)e^{ik_{z}z},~\left(\begin{array}{c}\Psi^{*}_{4}(r)e^{-i(m+2)\varphi}\\\Psi^{*}_{3}(r)e^{-i(m+1)\varphi}\\\Psi^{*}_{2}(r)e^{-im\varphi}\\\Psi^{*}_{1}(r)e^{-i(m-1)\varphi}\end{array}\right)e^{ik_{z}z}.\label{Eq_Kramer}
\end{equation}
One can check that at the site $k_{z}=0$, these two states are related to each other via time-reversal transformation. Also, one state of Eq.~(\ref{Eq_Kramer}) has total angular momentum $F_{z}=m+1/2$, and the other state of Eq.~(\ref{Eq_Kramer}) has total angular momentum $F_{z}=-(m+1/2)$. In other words, if we obtain one eigenfunction, the other degenerate counterpart is no need to calculate, they are related via the above relationship. This result is very useful in the following perturbation calculations.

\section{\label{Sec_IV}1D hole gas in a longitudinal magnetic field}
We now consider the effects of an external magnetic field which is applied along the wire direction, i.e., the $z$ direction $\textbf{B}=(0,0,B)$. Also, here we choose a symmetrical gauge for the vector potential $\textbf{A}=(-\frac{1}{2}By,\frac{1}{2}Bx,0)$. The Hamiltonian of the hole confined in the nanowire reads
\begin{eqnarray}
H&=&\frac{1}{2m_{e}}\left\{\left(\gamma_{1}+\frac{5}{2}\gamma_{s}\right){\bf p}'^{2}-2\gamma_{s}({\bf p}'\cdot\textbf{J})^{2}\right\}\nonumber\\
&&+2\kappa\mu_{B}BJ_{z}+V(r),\label{Eq_Hamiltonian_m}
\end{eqnarray}
where ${\bf p}'=\textbf{p}+e\textbf{A}$ and $\kappa=3.41$ is the Luttinger magnetic constant for semiconductor Ge~\cite{PhysRevB.4.3460}. We note that here we only consider a weak magnetic field, which can be treated perturbatively in the following calculations. In the presence of the magnetic field, the hole Hamiltonian (\ref{Eq_Hamiltonian_m}) is no longer time-reversal invariant. It is expected that the spin degeneracy in the subbands shown in Fig.~\ref{Fig_subband} would be lifted by this external magnetic field. We write the Hamiltonian (\ref{Eq_Hamiltonian_m}) in perturbative series with respect to the magnetic field $B$, and only keep those terms would lift the spin degeneracy in the subband dispersions. The Hamiltonian (\ref{Eq_Hamiltonian_m}) can be written as [$\mu_{B}=e\hbar/(2m_{e})$]
\begin{equation}
H=H_{0}+H^{(p_{1})}+H^{(p_{2})}+H^{(p_{3})},
\end{equation}
where $H^{(p_{1})}=2\kappa\mu_{B}BJ_{z}$ is the Zeeman term, and $H^{(p_{2})}$ and $H^{(p_{3})}$ come from the orbital effects of the magnetic field with the following expressions
\begin{widetext}
\begin{eqnarray}
H^{(p_{2})}&=&-i\mu_{B}B\partial_{\varphi}\left(\begin{array}{cccc}\gamma_{1}+\gamma_{s}&0&0&0\\0&\gamma_{1}-\gamma_{s}&0&0\\0&0&\gamma_{1}-\gamma_{s}&0\\0&0&0&\gamma_{1}+\gamma_{s}\end{array}\right),\nonumber\\
H^{(p_{3})}&=&-\sqrt{3}\gamma_{s}\mu_{B}B\left(\begin{array}{cccc}0&-ik_{z}re^{-i\varphi}&e^{-2i\varphi}(-r\partial_{r}+i\partial_{\varphi})&0\\ik_{z}re^{-i\varphi}&0&0&e^{-2i\varphi}(-r\partial_{r}+i\partial_{\varphi})\\e^{2i\varphi}(r\partial_{r}+i\partial_{\varphi})&0&0&ik_{z}re^{-i\varphi}\\0&e^{2i\varphi}(r\partial_{r}+i\partial_{\varphi})&-ik_{z}re^{-i\varphi}&0\end{array}\right).
\end{eqnarray}
\end{widetext}
Both the eigenvalues and the corresponding eigenfunctions of $H_{0}$ have been given in Sec.~\ref{Sec_III}. Thus, we just need to calculate the subband splitting using the degenerate perturbation theory. The $g$-factor is an important physical  parameter for the hole~\cite{PhysRevB.78.033307}, its anisotropy and tunability have been studied extensively~\cite{PhysRevLett.85.4574,Semina:2015aa,PhysRevB.93.121408,PhysRevB.84.075343,PhysRevResearch.3.013081,Zhang:2021wj}. Once the spin splitting is obtained, we can obtain the effective subband $g$-factor of the hole gas defined as
\begin{equation}
g_{l}\equiv\frac{E(k_{z},\Uparrow)-E(k_{z},\Downarrow)}{\mu_{B}B}=g_{l1}+g_{l2}+g_{l3},
\end{equation} 
where $g_{l1,2,3}$ are the $g$-factor components contributed by the perturbations $H^{(p_{1,2,3})}$ respectively. In most cases, the physical properties of the 1D hole gas are determined by its low-energy subband dispersions. Here, we only focus on the $g$-factor of the low-energy subband dispersions, which are given by total angular momentum $|F_{z}|=1/2$ (see Fig.~\ref{Fig_subband}). The explicit expressions of the $g$-factor components of the these low-energy subbands read
\begin{eqnarray}
g_{l1}&=&8\pi\kappa\int^{R}_{0}drr\Big(\frac{3}{2}|\Psi_{1}(r)|^{2}+\frac{1}{2}|\Psi_{2}(r)|^{2}\nonumber\\
&&~~~~~~-\frac{1}{2}|\Psi_{3}(r)|^{2}-\frac{3}{2}|\Psi_{4}(r)|^{2}\Big),\nonumber\\
g_{l2}&=&4\pi\int^{R}_{0}drr\Big(-(\gamma_{1}+\gamma_{s})|\Psi_{1}(r)|^{2}+(\gamma_{1}-\gamma_{s})|\Psi_{3}(r)|^{2}\nonumber\\
&&~~~~~~+2(\gamma_{1}+\gamma_{s})|\Psi_{4}(r)|^{2}\Big),\nonumber\\
g_{l3}&=&8\sqrt{3}\pi\gamma_{s}{\rm Re}\Bigg\{\int^{R}_{0}drr\Big(\Psi^{*}_{1}(r)(r\partial_{r}+1)\Psi_{3}(r)\nonumber\\
&&+\Psi^{*}_{2}(r)(r\partial_{r}+2)\Psi_{4}(r)\Big)\Bigg\}\nonumber\\
&&+8\sqrt{3}\pi\gamma_{s}k_{z}{\rm Im}\Bigg\{\int^{R}_{0}drr^{2}\big(\Psi^{*}_{3}(r)\Psi_{4}(r)\nonumber\\
&&-\Psi^{*}_{1}(r)\Psi_{2}(r)\big)\Bigg\}.\nonumber\\
\end{eqnarray}
where ${\rm Re/Im}\{\cdots\}$ means taking the real/imaginary part of its argument.

\begin{figure}
\includegraphics{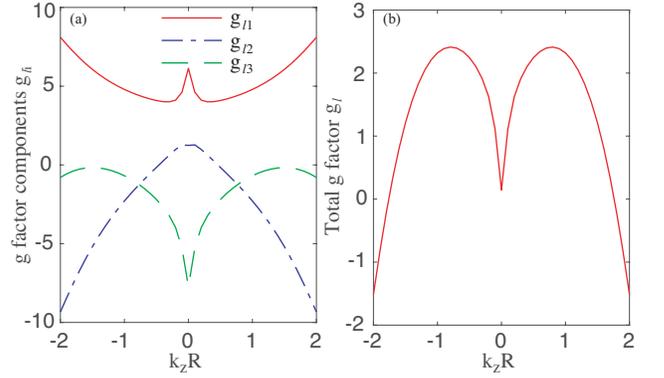}
\caption{\label{Fig_Gfacor_L}The dependence of the effective $g$-factor of the first lowest subband shown in Fig.~\ref{Fig_subband}(a) on the wave vector $k_{z}$. The magnetic field is applied along the nanowire. (a) The results for the components of the $g$-factor $g_{l1,2,3}$. (b) The result for the total $g$-factor $g_{l}$.}
\end{figure}

In Fig.~\ref{Fig_Gfacor_L}, we show the $k_{z}$-dependence of the effective $g$-factor of the lowest subband of the hole gas shown in Fig.~\ref{Fig_subband}(a). The results for the three components $g_{l1,2,3,}$ of the total $g$-factor are shown in Fig.~\ref{Fig_Gfacor_L}(a), and the result for the total $g$-factor $g_{l}$ is shown in Fig.~\ref{Fig_Gfacor_L}(b). As can be seen clearly from Fig.~\ref{Fig_Gfacor_L}(a), the contributions from the orbital terms $H^{(p_{2,3})}$ to the $g$-factor are as important as that from the Zeeman term $H^{(p_{1})}$. In other words, when we are interested in the magnetic properties of the hole, we can not neglect the orbital terms $H^{(p_{2,3})}$, although the radius $R$ of the nanowire can be made very small. Also, at the center of the $k_{z}$ space, i.e., $k_{z}R=0$, the total $g$-factor of the lowest subband is $g_{l}\approx0.14$, in agreement with Ref.~\cite{PhysRevB.84.195314}. At the band minimum, i.e., $|k_{z}R|\approx0.5$, the total $g$-factor is $g_{l}\approx2.24$.

\begin{figure}
\includegraphics{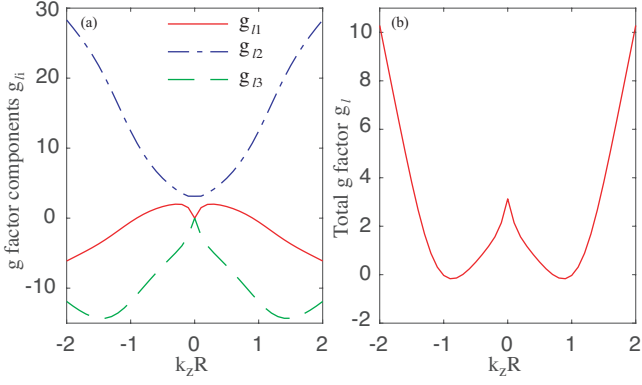}
\caption{\label{Fig_Gfacor_L2}The dependence of the effective $g$-factor of the second lowest subband shown in Fig.~\ref{Fig_subband}(a) on the wave vector $k_{z}$. The magnetic field is applied along the nanowire. (a) The results for the components of the $g$-factor $g_{l1,2,3}$. (b) The result for the total $g$-factor $g_{l}$.}
\end{figure}

In Fig.~\ref{Fig_Gfacor_L2}, we show the $k_{z}$-dependence of the effective g-factor of the second lowest subband of the hole gas shown in Fig.~\ref{Fig_subband}(a). The results for the three components $g_{l1,2,3,}$ of the total $g$-factor are shown in Fig.~\ref{Fig_Gfacor_L2}(a), and the result for the total $g$-factor $g_{l}$ is shown in Fig.~\ref{Fig_Gfacor_L2}(b). The contributions from the orbital terms $H^{(p_{2,3})}$ to the $g$-factor are still can not be neglected. Let us focus on the total $g$-factor $g_{l}$ near the site $k_{z}R=0$. Recall that there is an energy anticrossing between the first and the second lowest subbands shown in Fig.~\ref{Fig_subband}(a) at this special site. Hence, it is not strange to see a dip near this site in the $k_{z}$-dependence of the first lowest subband $g$-factor $g_{l}$ [see Fig.~\ref{Fig_Gfacor_L}(b)], and a peak near this site in the $k_{z}$-dependence of the second lowest subband $g$-factor $g_{l}$ [see Fig.~\ref{Fig_Gfacor_L2}(b)]. At $k_{z}R=0$, the total $g$-factor of the second lowest subband is $g_{l}\approx3.13$.

\section{\label{Sec_V}1D hole gas in a transverse magnetic field}
We now consider the situation where the magnetic field is applied perpendicular to nanowire, i.e., ${\bf B}=(B,0,0)$. The vector potential can be conveniently chosen as ${\bf A}=(0,0,By)$. It is easy to check that $\nabla\times{\bf A}={\bf B}$. The Hamiltonian of the hole confined in the nanowire now reads
\begin{eqnarray}
H&=&\frac{1}{2m_{e}}\left\{\left(\gamma_{1}+\frac{5}{2}\gamma_{s}\right){\bf p}'^{2}-2\gamma_{s}({\bf p}'\cdot\textbf{J})^{2}\right\}\nonumber\\
&&+2\kappa\mu_{B}BJ_{x}+V(r),\label{Eq_Hamiltonian_m2}
\end{eqnarray}
where ${\bf p}'=\textbf{p}+e\textbf{A}$. The Hamiltonian written in perturbation series with respect to the magnetic field $B$ reads (only those terms would lift the spin degeneracy are retained)
\begin{equation}
H=H_{0}+H^{(p_{1})}+H^{(p_{2})}+H^{(p_{3})},
\end{equation}
where $H^{(p_{1})}=2\kappa\mu_{B}BJ_{x}$ is the Zeeman term, and $H^{(p_{2})}$ comes from the orbital effect of the magnetic field
\begin{widetext}
\begin{eqnarray}
H^{(p_{2})}&=&2\sqrt{3}\gamma_{s}\mu_{B}B\left(\begin{array}{cccc}0&e^{-i\varphi}\sin\varphi(ir\partial_{r}+\partial_{\varphi})&0&0\\e^{i\varphi}\sin\varphi(ir\partial_{r}-\partial_{\varphi})&0&0&0\\0&0&0&e^{-i\varphi}\sin\varphi(-ir\partial_{r}-\partial_{\varphi})\\0&0&e^{i\varphi}\sin\varphi(-ir\partial_{r}+\partial_{\varphi})&0\end{array}\right),\nonumber\\
H^{(p_{3})}&=&2\mu_{B}Bk_{z}r\sin\varphi\left(\begin{array}{cccc}\gamma_{1}-2\gamma_{s}&0&0&0\\0&\gamma_{1}+2\gamma_{s}&0&0\\0&0&\gamma_{1}+2\gamma_{s}&0\\0&0&0&\gamma_{1}-2\gamma_{s}\end{array}\right).
\end{eqnarray}
\end{widetext}
The procedures are just the same as that for the longitudinal magnetic field case, we can calculate the subband splitting and define the effective subband $g$-factor as
\begin{equation}
g_{t}\equiv\frac{E(k_{z},\Uparrow)-E(k_{z},\Downarrow)}{\mu_{B}B}=\left|g_{t1}+g_{t2}+g_{t3}\right|,
\end{equation} 
where $g_{t1,2}$ are the $g$-factor components contributed by the perturbations $H^{(p_{1,2})}$ respectively, and $|\cdots|$ means taking the absolute value of its argument. The explicit expressions of the $g$-factor components of the subbands with total angular momentum $|F_{z}|=1/2$ read
\begin{eqnarray}
g_{t1}&=&8\pi\kappa\int^{R}_{0}drr\Big(\sqrt{3}\Psi_{1}(r)\Psi_{3}(r)+\Psi_{2}(r)\Psi_{2}(r)\Big),\nonumber\\
g_{t2}&=&8\sqrt{3}\pi\gamma_{s}\int^{R}_{0}drr\Big(\Psi_{1}(r)r\partial_{r}\Psi_{3}(r)\nonumber\\
&&+\Psi_{2}(r)(r\partial_{r}+2)\Psi_{4}(r)\Big),\nonumber\\
g_{t3}&=&8i\pi\,k_{z}\int^{R}_{0}drr^{2}\Big((\gamma_{1}+2\gamma_{s})\Psi_{2}(r)\Psi_{3}(r)\nonumber\\
&&+(\gamma_{1}-2\gamma_{s})\Psi_{1}(r)\Psi_{4}(r)\Big).
\end{eqnarray}
Note that $g_{t1,2}$ are complex numbers in general.

\begin{figure}
\includegraphics{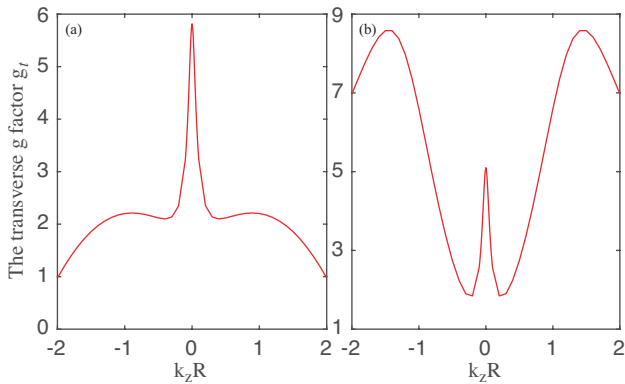}
\caption{\label{Fig_Gfacor_T}(a) The dependence of the effective $g$-factor of the first lowest subband shown in Fig.~\ref{Fig_subband}(a) on the wave vector $k_{z}$.  (b) The dependence of the effective $g$-factor of the second lowest subband shown in Fig.~\ref{Fig_subband}(a) on the wave vector $k_{z}$. The magnetic field is applied perpendicular to nanowire.}
\end{figure}


We show in Fig.~\ref{Fig_Gfacor_T} the $k_{z}$-dependence of the effective $g$-factors $g_{t}$ of the hole gas. The result for the first lowest subband given in Fig.~\ref{Fig_subband}(a) is shown in Fig.~\ref{Fig_Gfacor_T}(a), and the result for the second lowest subband given in Fig.~\ref{Fig_subband}(a) is shown in Fig.~\ref{Fig_Gfacor_T}(b). The orbital effect of the magnetic field $H^{(p_{2})}$ is still as important as the Zeeman term $H^{(p_{1})}$ in contributing to the effective hole $g$-factor. Due to the reason an energy anticrossing is located at the site $k_{z}R=0$ in the subband dispersions, there are sharp peaks in both Figs.~\ref{Fig_Gfacor_T}(a) and~\ref{Fig_Gfacor_T}(b) near this site. Also, at $k_{z}R=0$, the total $g$-factor of the lowest subband is $g_{t}\approx5.82$, in agreement with Ref.~\cite{PhysRevB.84.195314}, and the total $g$-factor of the second lowest subband is $g_{t}\approx5.10$. At the band minimum, i.e., $|k_{z}R|\approx0.5$, the total $g$-factor is $g_{t}\approx2.12$. Note that our result consists with an experimentally measured $g$-factor $g^{*}\approx2$ reported in Ref.~\cite{PhysRevLett.101.186802}.

\section{\label{Sec_VI}Hole spin qubit in nanowire quantum dot}
Via placing the Ge nanowire on a series of metallic gates, a nanowire single or double quantum dot can be achieved experimentally~\cite{PhysRevLett.101.186802,PhysRevLett.112.216806}. When the characteristic length of the longitudinal confinement (along the $z$ direction) is much larger than the nanowire radius $R$, the kinetic energy of the hole in the nanowire quantum dot can be well represented by the lowest two subbands given in Fig.~\ref{Fig_subband}(a). Because of this complicated kinetic energy for the hole in 1D~\cite{PhysRevB.84.195314}, the theoretical consideration of the hole spin qubit in the nanowire quantum dot may be more complicated than that in the planar Ge quantum dot~\cite{PhysRevB.103.125201}. Although the longitudinal $g$-factor is very small at $k_{z}R=0$, as we have emphasized previously the band minimum is located nearly at $|k_{z}R|\approx0.5$, such that both the longitudinal and the transverse magnetic fields are viable to produce the Zeeman splitting for the hole spin. The large $k_{z}$ dependence of the $g$-factor may also introduce some troubles in the theoretical consideration of the hole spin qubit. Anyway, it is possible to achieve a strong Rabi frequency in the electric-dipole spin resonance of the hole spin qubit~\cite{Froning:2021aa,wang2020ultrafast}, because when the lowest two subbands are used to govern the kinetic energy of the hole~\cite{PhysRevB.84.195314}, the intrinsic hole spin-orbit coupling is indeed very large [see Fig.~\ref{Fig_subband}(a)].  Also, for potential hole spin dephasing coming from charge noise~\cite{Wang:2021wc}, we suggest to design a symmetrical quantum dot confining potential along the wire direction, because the interplay between the asymmetrical confining potential and the spin-orbit coupling can lead to a longitudinal spin-charge interaction~\cite{lirui2018a,Li2020charge}.

\section{\label{Sec_VII}Discussion and conclusions}
The coupling between the orbital and spin angular momentum has been shown to give rise to an enhanced effective $g$-factor in the higher subbands of the electron gas in semiconductor nanowire~\cite{PhysRevLett.119.037701}. It is possible this coupling also affects the effective hole spin $g$-factor. For the low-energy subbands of the hole gas given by total angular momentum $|F_{z}|=1/2$ studied here, because the orbital angular momentum is small, this coupling may give a small amendment to the effective hole $g$-factor.

In this paper, we follow the method introduced in the seminal paper (see Ref.~\cite{PhysRevB.42.3690}) to study the effective $g$-factor of a 1D hole gas.  We obtain exactly both the subbands and the corresponding subband wave-functions of the hole gas in a cylindrical Ge nanowire. The subbands are two-fold degenerate, and the band minimum is not at the center of the $k_{z}$ space. A  simple relation between the degenerate subband wave-functions is obtained by exploring the time-reversal symmetry and the spin-rotation symmetry of the model. We then consider the effects of magnetic fields applied to the 1D hole gas both longitudinally and transversely. We evaluate the effective $g$-factor in the induced low-energy subbands. There is an energy anticrossing at the site $k_{z}R=0$, such that sharp dip and sharp peak structures appear near this site in the effective $g$-factor. The longitudinal $g$-factor $g_{l}$ is much less than the transverse $g$-factor $g_{t}$ at the site $k_{z}R=0$ for the lowest subband. While away from this site, e.g., at the band minimum $|k_{z}R|\approx0.5$, $g_{l}$ can be comparable to or even larger than $g_{t}$.

\section*{Acknowledgements}
This work is supported by the National Natural Science Foundation of China Grant No.~11404020, the Project from the Department of Education of Hebei Province Grant No. QN2019057, and the Starting up Foundation from Yanshan University Grant No. BL18043.

\appendix
\section{\label{Appendix_A}The matrix forms of $J_{x,y,z}$ and $\Gamma_{1,3}$}
In this papers, the detailed forms of the operators $J_{x,y,z}$ and $\Gamma_{1,3}$ are as follows:
\begin{equation}
J_{x}=\left(\begin{array}{cccc}
		0&\frac{\sqrt{3}}{2}&0&0\\
		\frac{\sqrt{3}}{2}&0&1&0\\
		0&1&0&\frac{\sqrt{3}}{2}\\
		0&0&\frac{\sqrt{3}}{2}&0
	\end{array}\right),
\end{equation}
\begin{equation}
J_{y}=\left(\begin{array}{cccc}
	0&-i\frac{\sqrt{3}}{2}&0&0\\
	i\frac{\sqrt{3}}{2}&0&-i&0\\
	0&i&0&-i\frac{\sqrt{3}}{2}\\
	0&0&i\frac{\sqrt{3}}{2}&0
\end{array}\right),
\end{equation}
\begin{equation}
J_{z}=\left(\begin{array}{cccc}
	\frac{3}{2}&0&0&0\\
	0&\frac{1}{2}&0&0\\
	0&0&-\frac{1}{2}&0\\
	0&0&0&-\frac{3}{2}
\end{array}
\right),
\end{equation}
\begin{equation}
\Gamma_{1}=\left(\begin{array}{cccc}0&0&i&0\\0&0&0&i\\-i&0&0&0\\0&-i&0&0\end{array}\right),~~~\Gamma_{3}=\left(\begin{array}{cccc}0&-i&0&0\\i&0&0&0\\0&0&0&i\\0&0&-i&0\end{array}\right).
\end{equation}

\section{\label{Appendix_B}The transcendental equation determining the subbands}
Via solving the eigenvalue equation of the Luttinger-Kohn Hamiltonian (\ref{Eq_LK}) in the cylindrical coordinate system, we have 
one branch of the bulk dispersion relation
\begin{equation}
E=(\gamma_{1}+2\gamma_{s})\frac{\hbar^{2}(\mu^{2}+k^{2}_{z})}{2m_{e}}.\label{Eq_dispersion1}
\end{equation}
The corresponding bulk wave-functions are~\cite{PhysRevB.42.3690}
\begin{equation}
\left(\begin{array}{c}
\frac{2ik_{z}}{\mu}J_{m-1}(\mu\,r)e^{i(m-1)\varphi}\\
\frac{4k^{2}_{z}+\mu^{2}}{\sqrt{3}\mu^{2}}J_{m}(\mu\,r)e^{im\varphi}\\
0\\
J_{m+2}(\mu\,r)e^{i(m+2)\varphi}
\end{array}\right)e^{ik_{z}z},
\end{equation}
and
\begin{equation}
\left(\begin{array}{c}
\sqrt{3}J_{m-1}(\mu\,r)e^{i(m-1)\varphi}\\
-\frac{2ik_{z}}{\mu}J_{m}(\mu\,r)e^{im\varphi}\\
J_{m+1}(\mu\,r)e^{i(m+1)\varphi}\\
0
\end{array}\right)e^{ik_{z}z}.
\end{equation}
Here $J_{m}(\mu\,r)$ is the $m$-order Bessel function. Another branch of the bulk dispersion relation reads
\begin{equation}
E=(\gamma_{1}-2\gamma_{s})\frac{\hbar^{2}(\mu^{2}+k^{2}_{z})}{2m_{e}}.\label{Eq_dispersion2}
\end{equation}
The corresponding bulk wave-functions are~\cite{PhysRevB.42.3690}
\begin{equation}
\left(\begin{array}{c}
\frac{2ik_{z}}{\mu}J_{m-1}(\mu\,r)e^{i(m-1)\varphi}\\
-\sqrt{3}J_{m}(\mu\,r)e^{im\varphi}\\
0\\
J_{m+2}(\mu\,r)e^{i(m+2)\varphi}
\end{array}\right)e^{ik_{z}z},
\end{equation}
and
\begin{equation}
\left(\begin{array}{c}
-\frac{4k^{2}_{z}+\mu^{2}}{\sqrt{3}\mu^{2}}J_{m-1}(\mu\,r)e^{i(m-1)\varphi}\\
-\frac{2ik_{z}}{\mu}J_{m}(\mu\,r)e^{im\varphi}\\
J_{m+1}(\mu\,r)e^{i(m+1)\varphi}\\
0
\end{array}\right)e^{ik_{z}z}. 
\end{equation}
The eigenfunction of (\ref{Eq_model}) can be written as Eq.~(\ref{eq_wavefunction}).
The hard-wall boundary condition $\Psi(R,\varphi,z)=0$ gives rise to
\begin{widetext}
\begin{equation}
\left(\begin{array}{cccc}
\frac{2ik_{z}}{\mu_{1}}J_{m-1}(\mu_{1}\,R)&\sqrt{3}J_{m-1}(\mu_{1}\,R)&\frac{2ik_{z}}{\mu_{2}}J_{m-1}(\mu_{2}\,R)&-\frac{4k^{2}_{z}+\mu^{2}_{2}}{\sqrt{3}\mu^{2}_{2}}J_{m-1}(\mu_{2}\,R)\\
\frac{4k^{2}_{z}+\mu^{2}_{1}}{\sqrt{3}\mu^{2}_{1}}J_{m}(\mu_{1}\,R)&-\frac{2ik_{z}}{\mu_{1}}J_{m}(\mu_{1}\,R)&-\sqrt{3}J_{m}(\mu_{2}\,R)&-\frac{2ik_{z}}{\mu_{2}}J_{m}(\mu_{2}\,R)\\
0&J_{m+1}(\mu_{1}\,R)&0&J_{m+1}(\mu_{2}\,R)\\
J_{m+2}(\mu_{1}\,R)&0&J_{m+2}(\mu_{2}\,R)&0
\end{array}\right)\cdot\left(\begin{array}{c}
c_{1}\\
c_{2}\\
c_{3}\\
c_{4}
\end{array}\right)=0.\label{Eq_coeffeicents}
\end{equation}
We introduce dimensionless parameters
\begin{eqnarray}
\varepsilon&=&\frac{E}{\hbar^{2}/(m_{e}R^{2})},\nonumber\\
k'_{z}&=&k_{z}R,\nonumber\\
\mu'_{1}&=&\mu_{1}R=\sqrt{\frac{2\varepsilon}{(\gamma_{1}+2\gamma_{s})}-k'^{2}_{z}},\nonumber\\
\mu'_{2}&=&\mu_{2}R=\sqrt{\frac{2\varepsilon}{(\gamma_{1}-2\gamma_{s})}-k'^{2}_{z}}.\label{eq_dimensionless}
\end{eqnarray}
The transcendental equation determining the energy spectrum of the hole in the cylindrical wire reads
\begin{equation}
f_{m}(\varepsilon)\equiv{\rm det}\left(\begin{array}{cccc}
\frac{2ik'_{z}}{\mu'_{1}}J_{m-1}(\mu'_{1})&\sqrt{3}J_{m-1}(\mu'_{1})&\frac{2ik'_{z}}{\mu'_{2}}J_{m-1}(\mu'_{2})&-\frac{4k'^{2}_{z}+\mu'^{2}_{2}}{\sqrt{3}\mu'^{2}_{2}}J_{m-1}(\mu'_{2})\\
\frac{4k'^{2}_{z}+\mu'^{2}_{1}}{\sqrt{3}\mu'^{2}_{1}}J_{m}(\mu'_{1})&-\frac{2ik'_{z}}{\mu'_{1}}J_{m}(\mu'_{1})&-\sqrt{3}J_{m}(\mu'_{2})&-\frac{2ik'_{z}}{\mu'_{2}}J_{m}(\mu'_{2})\\
0&J_{m+1}(\mu'_{1})&0&J_{m+1}(\mu'_{2})\\
J_{m+2}(\mu'_{1})&0&J_{m+2}(\mu'_{2})&0
\end{array}\right)=0.\label{eq_transcendental}
\end{equation}
The above equation is an implicit equation of $\varepsilon$, the zeros of function $f_{m}(\varepsilon)$ give rise to the eigenvalues of Hamiltonian (\ref{Eq_model}). Because $f_{m}(\varepsilon)$ is a complex number for a general energy $\varepsilon$, we usually plot the modulus $|f_{m}(\varepsilon)|$ as a function of $\varepsilon$ (see Fig.~\ref{fig_trascen}). We can see from the figure, several zeros of $f_{0}(\varepsilon)$ at the given wave vector $k_{z}R=0.5$ can indeed be found. We also note that the first zero of $f_{0}(\varepsilon)$ shown in Fig.~\ref{fig_trascen} gives a spurious energy eigenvalue of Hamiltonian (\ref{Eq_model}). This is because when $\varepsilon=(\gamma_{1}-2\gamma_{s})k^{2}_{z}R^{2}/2$, it follows that $\mu'_{2}=0$ [see Eq.~(\ref{eq_dimensionless})], such that the determinant given in Eq.~(\ref{eq_transcendental}) would equal to zero naturally.
\begin{figure}
\includegraphics{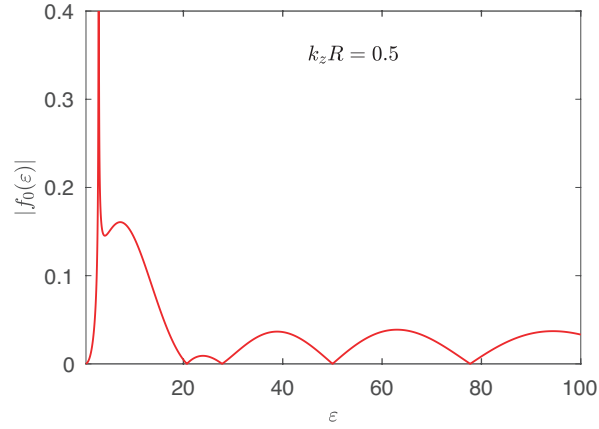}
\caption{\label{fig_trascen}The function $|f_{0}(\varepsilon)|$ is plotted in the interval $(\gamma_{1}-2\gamma_{s})k^{2}_{z}R^{2}/2\le\varepsilon\le100$. The zeros of function $|f_{0}(\varepsilon)|$, i.e., the sites $\varepsilon_{n}$ satisfy $|f_{0}(\varepsilon_{n})|=0$, are the eigenvalues of Hamiltonian (\ref{Eq_model}).  Note that here the first zero of $f_{0}(\varepsilon)$ is a spurious eigenvalue.} 
\end{figure}

\end{widetext}

\bibliographystyle{iopart-num}
\bibliography{Ref_Hole_spin}

\end{document}